\documentclass[12pt,preprint]{aastex}

\begin{document}

\title {CARBON DEFICIENCY IN EXTERNALLY-POLLUTED WHITE DWARFS: EVIDENCE FOR ACCRETION OF ASTEROIDS}

\author{M. Jura \\ Department of Physics and Astronomy and Center for Astrobiology  \\ University of California, 
 Los Angeles CA 90095-1562; jura@astro.ucla.edu}
\begin{abstract}
 Existing  determinations show that $n$(C)/$n$(Fe) is more than a  factor of 10  below solar in the atmospheres of three  white dwarfs  that appear to be externally-polluted.  These results are not easily explained if the stars have accreted interstellar matter, and we re-interpret these measurements as  evidence that these stars have accreted asteroids with a chondritic composition.   
  \end{abstract}
 \section{INTRODUCTION}
  In 
  white dwarfs with effective temperatures lower than about 20,000 K, it is predicted that heavy elements  settle below
  the photosphere in the star's strong gravitational field so the atmosphere is either largely hydrogen or largely helium  (Paquette et al. 1986, Chayer et al. 1995).  This expectation is supported by the determination that
  ${\sim}$75\% of these stars display upper limits to their calcium abundance that are as much as a  factor of 10$^{6}$ below the solar value (Zuckerman et al. 2003).  Currently, the only viable hypothesis to explain the observational result that ${\sim}$25\% of cool white dwarfs have detected metals is that these stars are externally-polluted. 
     Interstellar accretion (Paquette et al. 1986, Dupuis et al. 1993a,b, Koester \& Wilken 2006) is often thought to account for the metals within white dwarf atmospheres, but accretion from a tidally-disrupted minor body such as an analog to an asteroid or Kuiper  Belt Object naturally explains the infrared excesses and metal pollutions of  the hydrogen-rich white dwarfs G29-38, GD 362 and GD 56 (Jura 2003, Becklin et al. 2005, Kilic et al. 2005, 2006, Reach et al. 2005).    Here we re-interpret existing abundance determinations of three   helium-rich white dwarfs  that  previously were thought to have accreted interstellar matter  to argue, instead, that their measured low values of $n$(C)/$n$(Fe) support the hypothesis  that these stars have accreted circumstellar  matter with a chondritic composition.
    
 Since the diffusion times out of a white dwarf's atmosphere for different elements heavier than helium vary by less than a factor of 2.5 (Paquette et al. 1986, Dupuis et al. 1993a),  a  test of any accretion scenario is to compare the metals in the star's photosphere with the composition of the polluter.    Within meteorites of all kinds as well as the Earth's crust,  carbon is depleted relative to refractory species such as iron, magnesium and silicon  by at least a factor of 10 and often more.    Accretion of circumstellar matter with a  composition similar to rocky material in the inner solar system can lead to distinctively low values of $n$(C)/$n$(Fe) that cannot be naturally explained by accretion of interstellar matter.  On the basis of this argument,  a carbon deficiency in the photosphere of
the main sequence star J37 in the open cluster NGC 6633 has been attributed to the accretion of
material with a chondritic composition (Ashwell et al. 2005, Laws \& Gonzalez 2003).

        Carbon has been detected in some  helium-rich white dwarfs.  
 For the stars with $T_{eff}$ $<$ 13000 K, the convective zones are sufficiently deep that carbon may be
dredged up from the interior (Koester et al. 1982, Pelletier et al. 1986, Dupuis et al. 1993b,  MacDonald et al. 1998).      In three white dwarfs with $T_{eff}$ near 25000 K, Petitclerc et al. (2005) have reported an unexplained presence of enhanced carbon with, for example, $n$(C)/$n$(Si) $>$100.   Here we focus on 
stars with carbon deficiencies in the sense that $n(C)$ $<<$ $n(Fe)$.

The unique chemistry of carbon makes it a useful element for  distinguishing between circumstellar and interstellar models of atmospheric pollution.  In the early solar system,  carbon was probably largely concentrated in volatile molecules such as
CH$_{4}$ and CO (see, for example, Lodders 2003, 2004) and not incorporated into rocky bodies that formed in the inner solar system such as the Earth and asteroids (see also Simonelli et al. 1997, Gail 2002).  
In environments where carbon is more abundant than oxygen, carbon can condense
into refractory solids such as graphite, and this may have occurred during the formation of  some of the outer regions of  the solar system where water was depleted (see Krot et al. 2000) and also, perhaps, around ${\beta}$ Pic (Roberge et al. 2006).   In any case, meteorites, which are the best available measure of the composition of
asteroids, are deficient in carbon (Wasson \& Kallemeyn 1988).  In contrast, as discussed  below, carbon 
is abundant in both the gaseous and solid phase  in the interstellar medium.

In $\S2$ we discuss scenarios for interstellar accretion and describe why the alternative of accretion from circumstellar matter is preferable for many stars.  In $\S3$ we describe the scenario for  white dwarf pollutions by minor bodies such as asteroids.   In $\S4$ we discuss our results and
  in $\S5$ we report our conclusions. 

Below, we suggest that at least  7\% of  white dwarfs possess asteroid belts analogous to the solar system's.  However, two recent surveys  of main-sequence solar-type stars find  that only 1 star out of 69  (Bryden et al. 2006) or 1 star out of 41  (Beichman et al. 2006)  have warm dust that might be associated with asteroids.     While these surveys could not have detected an analog to the current zodiacal cloud in the solar system,  our own asteroid belt was probably much  more massive in the past, and  associated warm dust   could have produced  a significant infrared excess at
 ${\lambda}$ $<$ 30 ${\mu}$m (Gaidos 1999).  
 We show in an Appendix that when the effects of a stellar wind are included, the infrared excess from dust associated with asteroids is suppressed, and therefore,  both young and old solar-type main-sequence  stars may possess  undetected asteroid belts.  
  
\section{WHITE DWARF POLLUTION FROM THE INTERSTELLAR MEDIUM?}

The accretion of interstellar matter is a plausible source of the external-pollution of some white dwarfs.    Here, however, we argue that interstellar accretion  cannot explain the measured atmospheric abundances in the three helium-rich white dwarfs listed in Table 1.  These stars are taken from the study of Wolff et al. (2002) who report both their own results and summarize previous studies for the hydrogen, carbon, magnesium, silicon, calcium and iron abundances in 10 helium-rich white dwarfs.  The element with the smallest error in its abundance determinations is iron (Wolff et al. 2002), and here  we focus on abundances  relative to this element.
 For six  stars, the carbon abundance is not well enough measured to determine  the source of the material.   One star, L 119-34, exhibits $n$(C)/$n$(Fe) ${\approx}$ 100, but with very large errors; this star may have experienced dredge-up. The three stars with values of $n$(C)/$n$(Fe) more than a  factor of 10 below solar  are listed in Table 1.

  We show in Figure 1 a plot of $\log$ $n$(Mg)/$n${Fe), compared to $\log$ $n$(Si)/$n$(Fe),   for the Sun (Lodders 2003),
 CI chondrites (Lodders 2003),  L chondrites (Wasson \& Kallemeyn 1988), the Earth's crust (Ronov \& Yaroshevsky 1969) and the three white dwarfs listed in Table 1 (Wolff et al. 2002).    We only show results for L chondrites;
   the relative abundances in LL chondrites and H chondrites are within 30\% of those in L chondrites (Wasson \& Kallemeyn 1988).  Together, these three classes of chondrites comprise slightly more than 75\% of  all meteorite ``falls" (Table II-8 in Wasson 1974).  Also, we do not display the relative abundances for comet Halley discussed below since they are very close to the meteoritic values (Jessberger et al. 1988).

   We see from Figure 1  that relative silicon and magnesium abundances in H 2253+8023 and GD 40 are in agreement with the solar and meteoritic values while the relative abundances for Ross 640 are apparently discrepant.   However,  given all the uncertainties, the white dwarf abundances shown  are consistent with the picture that these stars accrete from the interstellar medium  (Wolff et al. 2002).   
   Figure 2 shows a plot  of $\log$ $n$(Ca)/$n$(Fe) compared to $\log$ $n$(Mg)/$n$(Fe).  Again, within the large  uncertainties,  the relative abundances in the atmospheres of the white dwarfs are consistent with  accretion of interstellar matter.
 
 We show in Figure 3 a comparison of $\log$ $n$(C)/$n$(Fe)  with $\log$ $n$(Mg)/$n$(Fe) for the same objects as shown in Figures 1 and 2.     The LL and H chondrites have values of $\log$ $n$(C)/$n$(Fe) similar to that of the L chondrites while all other chondrites have values of $\log$ $n$(C)/$n$(Fe) ranging between that of  the CI chondrites and that of the L chondrites.  Carbon composes  only 10$^{-4}$ of the mass of iron meteorites (Lewis \& Moore 1971).  
Figure 3 shows that all  three white dwarfs have carbon to iron abundance ratios at least a factor of 10 below solar;  a result that is 
  naturally understood if these stars have   accreted  asteroids with a chondritic composition.   

 In Figure 3 we also display the results for comet Halley (Jessberger et al.1988).  Its  relatively high carbon abundance, which is characteristic of many but not all comets (A'Hearn et al. 1995), means that  models for cometary accretion
 onto these white dwarfs (Alcock et al. 1986) are not supported for Halley analogs.  Analogs to Kuiper Belt Objects that are largely composed of ice or other volatiles are not likely to survive the star's evolution on the Asymptotic Giant Branch (Jura 2004).  However,  rocky cores of such objects might survive, and
 ultimately could accrete onto the star after it has evolved into a white dwarf.  We do not know if such hypothetical bodies are as carbon-deficient as required by the observed pollutions.
     
Is it possible to modify the interstellar accretion model to explain the results shown in Figure 3?  Because of their low hydrogen abundances, it is usually hypothesized that the externally-polluted helium-rich white dwarfs accrete grains without accreting gas;  the hydrogen accretion rate must be suppressed by as much as factor of 10$^{6}$ (Friedrich et al. 1999, 2004).  Is the interstellar carbon largely in the gas-phase and therefore not accreted? (see Frisch \& Slavin 2003, Kimura et al. 2003)
The interstellar gas-phase carbon abundance lies between 1.40 ${\pm}$ 0.20 ${\times}$ 10$^{-4}$ (Cardelli et al. 1996) and 1.61 ${\pm}$ 0.17 ${\times}$ 10$^{-4}$ (Sofia et al. 2004). The total solar carbon abundance, which may also be the interstellar abundance,  is given as 2.45 ${\pm}$ 0.26 ${\times}$ 10$^{-4}$ by Asplund et al. (2005), but might be as large as 3.3 ${\times}$ 10$^{-4}$ (Pinsonneault \& Delahaye 2006).  With these results, it appears that  carbon is appreciably distributed in both the gas phase and solid phase.  However, the true interstellar abundance may not be solar.  The abundance of carbon in early-type stars is inferred to be 1.95 ${\pm}$ 0.14 ${\times}$ 10$^{-4}$ by Nieva \& Przybilla (2006), but, in contrast,  Hempel \& Holweger (2003) find carbon  abundances in such stars often to be greater than solar. In any case,  models of interstellar grains require a large amount of carbon in the solid-phase.   Zubko et al. (2004) describe 15 possible interstellar grain models in their Table 5, and all  require at least 1.90 ${\times}$ 10$^{-4}$ carbon atoms in the grains.  It appears probable that interstellar carbon is very roughly equally divided between the gas and solid phases, and it is thus unlikely that almost all interstellar carbon is preferentially inhibited from accreting onto  white dwarfs.

\section{WHITE DWARF POLLUTION BY MINOR BODIES}

We now consider the implications of accreting an asteroid and make an order of magnitude estimate of the  a pollution event.  We therefore neglect  three possible complications which could quantitatively change our detailed results but probably do not lead to
 a qualitative change in our basic scenario for white dwarf accretion.  (1)  We assume that the outer convection zone is the region where the accreted material is well-mixed.  In main-sequence
 A-type stars, there may be an outer mixing zone which is substantially larger than the outer
 convection zone (Richer et al. 2000). (2)  We use  published calculations which assumed spherical symmetry for estimating the diffusion time scale.  At least for main-sequence stars, spherical symmetry may be an oversimplication (Vauclair 2004).  
 (3)  We implicitly assume a relative simple set of compositional layers in the helium rich white dwarf -- an outer envelope and an inner core.  Recent astroseismological studies of white dwarfs show that their interiors are more
 complex than previously assumed (Metcalfe 2005).  While these new insights have led to a better understanding of the interior structure of white dwarfs, they have not led to a change in the basic
paradigm that external pollution explains the presence of metals in the atmospheres of cool white dwarfs.
 
 Consider a pollution event.  
 While a minor body could directly impact a white dwarf, a more likely scenario is that an asteroid
  strays within the star's tidal radius.  In this case, a flat disk analogous to Saturn's rings could form
  (Jura 2003), and the matter then accretes from this ring onto the star.   We assume that the externally-donated material from the ring is quickly  and uniformly distributed throughout the star's outer convective envelope which has mass, $M(env)$. We assume that all the polluting elements linger in the atmosphere for the same characteristic settling time, and then the star again appears unpolluted.  The convective envelopes of helium-rich white dwarfs are much more massive than those of hydrogen-rich
  white dwarfs, and therefore the helium-rich white dwarfs require much greater amounts of pollution
  to produce a detectable abundance anomaly.  
  
  We now consider the individual stars in Table 1.    For GD 40 and HS 2253+8023 with $T_{eff}$ near 15,000 K we adopt $M(env)$ = 2 ${\times}$ 10$^{-6}$ M$_{\odot}$ while for Ross 640 with $T_{eff}$ =  8500 K we adopt $M(env)$ = 6 ${\times}$ 10$^{-6}$ M$_{\odot}$ (McDonald et al. 1998).  The mass of accreted iron in the outer  envelope, $M_{Fe}(env)$, is then derived from  the photospheric iron abundances relative to helium, the dominant constituent in the atmosphere, as reported by Wolff et al (2002). The inferred values of $M_{Fe}(env)$  are given in Table 1.  
  If $f_{Fe}$, the fraction of the accreted mass that is iron, is ${\sim}$0.2 as found in L chondrites (Wasson \& Kallemeyn 1988), then  the total inferred parent body mass is  $M_{Fe}(env)/f_{Fe}$. The masses of the polluting bodies that produced the  current accretion episodes are between  ${\sim}$7 ${\times}$ 10$^{-7}$ M$_{\oplus}$ and ${\sim}$3 ${\times}$ 10$^{-4}$ M$_{\oplus}$.   For comparison, the mass of Ceres, the largest asteroid in the solar system, is 1.6 ${\times}$ 10$^{-4}$ M$_{\oplus}$ (Michalak 2000).  
  
Above we have estimated the mass required to explain a single pollution event;  we now estimate the total amount of mass that is required in asteroid belts to explain the available abundance data
 for externally-polluted helium-rich white dwarfs as a class.     
 We use $f_{accrete}$ to denote the fraction of time that a star exhibits an accretion event and $f_{wd}$ to denote the fraction of white
dwarfs with asteroid belts.  We cannot independently  estimate $f_{accrete}$ and $f_{wd}$, but only their ratio.    
In the survey of 800 white dwarfs by  Koester et al. (2005) of which ${\sim}$20\% are likely to be helium-rich (Hansen 2004), there are 11 helium-rich white dwarfs with  $\log$ $n$(Ca)/$n$(He) $>$ -8.7, the calcium abundance in Ross 640.  Therefore for the ``typical" pollution event, we adopt  $(f_{accrete}/f_{wd})$ = 0.07.   
With $\log$ $n$(Ca)/$n$(He) = -6.1, the abundances in HS 2253+8023 are so high that for events with this extreme degree of pollution
we adopt $(f_{accrete}/f_{wd})$ = 0.01. 
We define $M_{accrete}$ as the total mass that a white dwarf accretes during all the pollution episodes that occur during its cooling time.
 Very approximately:
\begin{equation}
M_{accrete}\;{\approx}\,\frac{M_{Fe}(env)}{f_{Fe}}\,\frac{t_{cool}}{t_{settl}}\,\frac{f_{accrete}}{f_{wd}}
\end{equation}
where $t_{cool}$ denotes the white dwarf's cooling age taken from  Hansen (1999, 2004) and $t_{settl}$ the settling time out of the star's convective envelope.      We adopt $t_{settl}$ = 5 ${\times}$ 10$^{5}$ yr   (Paquettte et al. 1986, Dupuis et al. 1993a) for all elements at every effective temperature considered here.   
 We list in Table 1 our values of $M_{accrete}$ from equation (1); the results range from ${\sim}$7 ${\times}$ 10$^{-5}$ M$_{\oplus}$ to ${\sim}$10$^{-3}$ M$_{\oplus}$. 

\section{DISCUSSION}

Before a star becomes a white dwarf, it loses half or more of its main-sequence mass, and the orbits of all the planets and asteroids are altered.  Debes \& Sigurdsson (2002) have computed that there can be a resulting  
enhanced rate of asteroids being perturbed into high eccentricity orbits that lead them  near enough to the white
dwarf for catastrophic disruption to occur.  However, it is yet to be demonstrated that enough perturbations of this sort  can persist for    ${\sim}$1 Gyr as required to explain   polluted white dwarfs like Ross 640.   

 We have suggested that at least 7\% of 
 white dwarfs  have asteroid belts with total minimum masses of between ${\sim}$7 ${\times}$ 10$^{-5}$  M$_{\oplus}$ and ${\sim}$10$^{-3}$ M$_{\oplus}$.  Since   
  the current mass of the solar system's asteroid belt is ${\sim}$6 ${\times}$ 10$^{-4}$ M$_{\oplus}$ (Krasinsky et al. 2002), the external pollution of white dwarfs can be supplied by
  asteroid belts analogous to the solar system's.  In the Appendix, we argue that the current evidence that only ${\sim}$2\% of solar-type main-sequence stars exhibit evidence of asteroids does not exclude this proposed scenario.  
  
 Carbon  abundances are  useful for the purpose of distinguishing interstellar from circumstellar accretion.     While it is possible to measure values of $n$(C)/$n$(He) as low as  ${\sim}$10$^{-6}$ for relatively cool helium-rich white dwarfs from the optical C$_{2}$ bands (Koester et al. 1982), measurements or upper limits of  $n$(C)/$n$(He) as low as 10$^{-9}$ around the hotter white dwarfs    have required ultraviolet measurements from space-borne telescopes (Wegner \& Nelan 1987, Wolff et al. 2002).  

The diffusion time of heavy elements out of the atmosphere of a hydrogen-rich white dwarf is typically less than 1000 years.  Such polluted stars must be experiencing ongoing accretion.  However,
the helium-rich stars have much thicker outer convective envelopes, and their diffusion times are correspondingly longer.  Polluted helium-rich white dwarfs may not currently exhibit an infrared excess.  

A difficulty with the model for accretion of tidally-disrupted asteroids is that some externally-polluted hydrogen-rich white dwarfs do not display any evidence of circumstellar dust (Kilic et al. 2006).  It is possible that the dust rings around G29-38 and GD 362 are particularly opaque, analogous to Saturn's rings, while some white dwarfs have more transparent dust rings  or their disks are largely gaseous.   In any case, 
the hypothesis that white dwarfs accrete interstellar grains also leads to the expectation that
 these stars should  dispay an infrared excess, and therefore the absence of such an  excess  is not a strong argument in favor of the interstellar  accretion model.

\section{CONCLUSIONS}
The notably low  carbon  to iron ratios  in the atmospheres of three white dwarfs are naturally  explained if these stars have accreted  carbon-poor asteroids with masses between  7 ${\times}$ 10$^{-7}$ M$_{\oplus}$ and  3 ${\times}$ 10$^{-4}$ M$_{\oplus}$.    The inferred existence of extrasolar carbon-deficient minor bodies  suggests that chemical and dynamical scenarios for the formation and evolution of  rocky planetesimals in the inner solar system can be generalized.

I thank B. Hansen, D. Koester, J. Wasson  and B. Zuckerman for valuable conversations.  This work has been partly supported by NASA.  
 
 \appendix
 \begin{center}
 {\bf APPENDIX}
 \end{center}
 \renewcommand{\theequation}{A\arabic{equation}}
 \setcounter{equation}{0}
Is there  enough mass in  asteroid belts around main-sequence stars to account for 
white dwarf pollutions?  Paralleling the discussion  by Plavchan et al. (2005) for main-sequence
M-type stars, we argue that when the effects of stellar winds are included, infrared excesses from analogs  to the Sun's asteroid belt are difficult to detect even for young main-sequence stars.  

 Poynting-Robertson drag controls the grain lifetime in the solar system (Burns et al. 1979). However,
when a star of luminosity, $L_{*}$, has a  mass loss rate, ${\dot M}_{wind}$, sufficiently large such that ${\dot M}_{wind}\,c^{2}$ $>$ $L_{*}$, as often occurs around stars younger than the Sun, then drag from this wind controls the particle lifetime (Jura 2004, Plavchan et al. 2005).  The lifetime, $t_{gr}$, of a spherical grain of radius $b$  composed of material with density, ${\rho}_{dust}$  in a circular orbit of radius $D$  exposed to a stellar wind with rate ${\dot M}_{wind}$  in the drag-dominated regime, is (Plavchan et al. 2005):
\begin{equation}
t_{gr}\;{\approx}\;\frac{4\,{\pi}\,b\,{\rho}_{dust}\,D^{2}}{3\,{\dot M}_{wind}}
\end{equation}
With $b$ = 30 ${\mu}$m as representative of particles in the zodiacal cloud of the solar system (Fixsen \& Dwek 2002), $D$ = 3 AU, ${\rho}_{dust }$ = 3 g cm$^{-3}$ and ${\dot M}_{wind}$ ${\sim}$ 3 ${\times}$ 10$^{14}$ g s$^{-1}$ as characteristic of a vigorous stellar wind (see below), then $t_{gr}$ ${\approx}$ 8000 yr.  Therefore, the lifetime of dust particles is short compared to the main-sequence age, and the systems quickly relax to a steady state.  In this case, the luminosity of the infrared excess, $L_{IR}$, compared to the stellar luminosity, $L_{*}$, is (Jura 2004):
\begin{equation}
L_{IR}\;{\approx}\;\frac{{\dot M_{dust}}}{2\,{\dot M_{wind}}}\;L_{*}\;\ln\left(\frac{D_{init}}{D_{final}}\right)
\end{equation}
where $D_{init}$ and $D_{final}$ denote the initial and final distances of a dust grain from the host star and ${\dot M}_{dust}$  denotes the dust production rate by asteroidal grinding. 
Averaging over time and for the moment setting the numerical coefficient equal to unity, we find from (A2) that:
\begin{equation}
<\frac{L_{IR}}{L_{*}}>\;{\approx}\;\frac{{\Delta}M_{dust}}{{\Delta}M_{wind}}
\end{equation}
where ${\Delta}M_{dust}$ and ${\Delta}M_{wind}$ denote the total mass in dust produced by asteroids and the total mass lost by a stellar wind.  Equation (A3) shows that the fractional infrared excess depends upon the mass in dust created from asteroids compared to the mass lost from the host star.  

 A promising model for the origin and evolution of the solar system's asteroids is that the region between Mars and Jupiter initially had ${\sim}$1 M$_{\oplus}$ of planetesimals.  After Jupiter formed, this population was dynamically disrupted until perhaps only
 0.01 M$_{\oplus}$ of material remained after the initial 100 Myr of the solar system  (Petit et al. 2001, Bottke et al. 2005).   Subsequently, this population of asteroids was slowly reduced by mutual collisions which ultimately lead to dust and produced an infrared excess.  If the Sun loses 0.01 M$_{\odot}$ in its wind (see below), then we expect a time averaged value of $L_{IR}/L_{*}$ of ${\sim}$ 3 ${\times}$ 10$^{-6}$ which, as illustrated below, is difficult to detect.  If the dust production rate is stochastic as has been suggested for the history of the asteroid belt (Durda \& Dermott 1997) and
 for dust production around main-sequence A-type stars (Rieke et al. 2005), then there can be phases
 when the infrared excess is detectable.  

Above, we have discussed some general features of the detectability of asteroids; we now present a specific illustration using conservative estimates for poorly-known  quantities.  Since dust is produced by collisions, we expect that ${\dot M}_{dust}$ =
$C\,M_{PB}^{2}$ where $M_{PB}$ is the mass of the asteroid belt and $C$ is a complicated function determined by the composition, size distribution and orbits of the asteroids.   Instead of evaluating $C$ from first principles, we adopt
an empirical approach and implicitly derive  $C$ from the observed rate of dust production in  the solar system.  This method can explain why we do not frequently detect analogs to our own
asteroid belt; it does not  explore the physics of asteroid formation and evolution.

 In a simple model  $C$ is constant with time (Dominik \& Decin 2003) and:
\begin{equation}
{\dot M}_{dust}(t)\;=\;{\dot M}_{dust}(t_{Now})\,\left(\frac{t_{Now}}{t}\right)^{2}
\end{equation}
where $t_{Now}$ denotes the current age of the Sun which we take to equal 4.6 Gyr.  Alternatively,  Gaidos (1999) has suggested that $C$ can be scaled from the history of cratering on
on the Moon.  In this case, 
\begin{equation}
{\dot M}_{dust}(t)\;=\;{\dot M}_{dust}(t_{Now})\,\left(1\,+\,{\beta}'\,\exp\left[\frac{t_{Now}\,-\,t}{{\tau}_{cr}}\right]\right)
\end{equation}
where ${\beta}'$ = 1.6 ${\times}$ 10$^{-10}$ and ${\tau}_{cr}$= 0.144 Gyr (Chyba 1991).  We show a comparison between the two production rates in Figure 4; ${\dot M}_{dust}$ was much larger in the past.    We do not show dust production rates for ages less than 100 Myr because these
extrapolations are especially uncertain.    Equations (A4) and (A5) integrate to yield total dust production  from $t$ = 100 Myr until now of ${\sim}$0.003 M$_{\oplus}$ and ${\sim}$0.01 M$_{\oplus}$, respectively,  for ${\dot M}_{dust}(t_{Now})$ = 3 ${\times}$ 10$^{6}$ g s$^{-1}$ (see below).  These integrated dust masses are comparable to  the very uncertain estimate of the total amount
of mass in asteroids that existed after the first 100 Myr of the solar system.  

To assess equation (A2), we need to estimate the spatial motion of the dust.  Because the initial and final 
locations  enter only logarithmically, 
we incorporate very simply the effects of particle collisions and orbiting planets by setting
$D_{init}$ = 3 $D_{final}$.  To find the specific luminosity, 	
we expect the  particles to drift inwards and establish a density variation that scales as $D^{-1}$ and therefore that $L_{IR,\,{\nu}}$ (erg s$^{-1}$ Hz$^{-1}$) varies as ${\nu}^{-1}$ over the spectral range where most of the emission is produced (Jura 2004, Waters et al. 1988).  In this case:
\begin{equation}
L_{IR,\,{\nu}}\;{\approx}\;\frac{{\dot M}_{dust}}{2\,{\dot M}_{wind}}\,\frac{L_{*}}{\nu}
\end{equation}
  Equation (A6) is valid between a minimum frequency, ${\nu}_{min}$, and a maximum frequency, ${\nu}_{max}$ given by ${\nu}_{min}$ ${\approx}$ $3\,k_{B}\,T_{min}/h$ and $3\,k_{B}\,T_{max}/h$, respectively where the minimum and maximum grain temperatures are determined by $D_{init}$ and $D_{final}$.     Adopting  $D_{init}$ = 3 AU, then a particle that behaves like a black boy  in orbit around the Sun has a temperature of 160 K implying that the  maximum wavelength at which we can use equation (A6)  is ${\sim}$30 ${\mu}$m.

To estimate the fractional infrared excess, $f_{ex}$,
we compare the emission of the circumstellar dust with  the long-wavelength emission from the star's atmosphere, $L_{*,\,{\nu}}$, which we take as:
\begin{equation}
L_{*,\,{\nu}}\;{\approx}\;4\,{\pi}\,R_{*}^{2}\,\frac{2\,{\pi}\,{\nu}^{2}\,k_{B}\,T_{*}\,k_{corr}}{c^{2}}
\end{equation}
 In equation (A7), we expect that the stellar atmosphere emits on the Rayleigh-Jeans slope but the absolute value of the emission scales from the star's effective temperature by the factor   $k_{corr}$.    For the Sun at 20 ${\mu}$m, we use the central surface brightness (Neckel 2000) and the limb darkening (Pierce 2000) to derive $k_{corr}$ = 0.84 which we adopt  here.
 
 The mass loss rate from solar-type main-sequence stars are measured for only a handful of stars and are   poorly known.    Sackmann \& Boothroyd (2003) have reviewed the available observations of other stars and also used constraints from helioseismology, the current solar lithium abundance, and the signature of the solar wind irradiation of the lunar surface to infer possible historical mass loss rates from the Sun. They consider 21 models with a total mass loss ranging from 0.01 M$_{\odot}$ to 0.07 M$_{\odot}$.  Here, we adopt their most conservative estimate that:
 \begin{equation}
{\dot M}_{wind}(t)\;=\;{\dot M}_{wind}(0)\,e^{\left(-\frac{t}{{\tau}_{wind}}\right)}
\end{equation}
where ${\dot M}_{wind}(0)$ = 8.3 ${\times}$ 10$^{14}$ g s$^{-1}$ and ${\tau}_{wind}$ = 0.755 Gyr. In this model, the Sun loses 1\% of its initial mass.  Wood et al. (2005) have recently proposed that
the mass loss rates are lower than previously thought among stars of high activity.  However, Chen et al. (2005) have found an anti-correlation between the X-ray luminosity and infrared excess at 24 ${\mu}$m of  solar-type main sequence stars with ages near ${\sim}$10 Myr which might be the result of strong winds  associated with X-ray activity also suppressing  the infrared emission.  Even though it is described as conservative, equation (A8) is highly uncertain.

 To estimate the infrared excess, we assume the dust production rate given by equation (A4), and then
using  (A6) - (A8),  we find:
\begin{equation}
f_{ex}(t\,,{\lambda})\;=\;\frac{L_{IR,\,{\nu}}}{L_{*,\,{\nu}}}\;=\;\left(\frac{{\dot M}_{dust}(t_{Now})}{{\dot M}_{wind}(0)}\right)\,\left(\frac{t_{Now}}{t}\right)^{2}\,\left(\frac{{\sigma}_{SB}\,T_{*}^{3}\,{\lambda}^{3}}{4\,{\pi}\,k_{corr}\,k_{B}\,c}\right)\,e^{\left(\frac{t}{{\tau}_{wind}}\right)} 
\end{equation}
where ${\sigma}_{SB}$ denotes the Stephan-Boltzmann constant.    Making the extreme assumption that  all the dust in the zodiacal clourd arises from asteroids, we adopt
 ${\dot M}_{dust}(t_{Now})$ = 3 ${\times}$ 10$^{6}$ g s$^{-1}$ (Fixsen \& Dwek 2002).  An equivalent to equation (A9) can be derived for the dust production rate given by equation (A5) instead of equation (A4).

  We show in Figure 5  plots 
of $f_{ex}$(24 ${\mu}$m)   to illustrate  the infrared excess that might be produced as a function of time.   We show results for 24 ${\mu}$m because the dust emission stands out best at the longest wavelengths, as can be seen from equation (A9).  Also, the ${\it Spitzer\; Space\; Telescope}$ can observe
sensitively  at this wavelength.  Finally,  we are interested in  asteroids cooler than 250 K, since any hotter asteroids  would probably be destroyed when the star evolves onto the asymptotic giant branch (see, for example, Rybicki \& Denis 2001).  Our display in Figure 5  overestimates $f_{ex}$(24 ${\mu}$m)  for stars older than 3 Gyr since Poynting-Robertson drag is not included.   Also,  we do not  show $f_{ex}$(24 ${\mu}$m)  for stars younger than 100 Myr because the extrapolations are unreliable. 

 We see in Figure 5 that  the infrared excess is  unobservable for ${\dot M}_{dust}$ given by equation (A4).    However, if the dust production rate follows equation (A5), then when the system is young, the excess may be detectable for a duration of  ${\sim}$200 Myr or about 4\% of the Sun's total current age. Thus even if equation (A5) correctly approximates ${\dot M}_{dust}$, we expect the detection of infrared excesses from asteroids to be rare.   The displays in Figure 5 are for a ``conservative" estimate of the stellar wind mass loss rate;
  $f_{ex}$(24 ${\mu}$m) would be lower if the wind is stronger than given by equation (A8).  
 
 It is possible that instead of a steady production of dust given by ${\dot M_{dust}}$ there are short, impulsive bursts  from destruction of large asteroids.  We now present a schematic description of how this might affect the observability of infrared excesses.  We denote the instantaneous dust formation rate  by ${\dot M_{inst}}$.   We assume that for times long compared to the particle decay time given by equation (A1) but short compared to the stellar evolution time that  $<{\dot M_{inst}}>$ equals ${\dot M_{dust}}$. We also assume that during a stochastic event when the particle decay time  is relatively short compared to the time that it takes for an asteroid to be fully pulverized,   the infrared excess  can be derived from the instantaneous dust decay rate and equation (A2). 
 
  Let $f_{det}({\lambda})$ denote the threshold for detecting an excess at wavelength ${\lambda}$.        If $p(t,\,{\lambda})$ denotes the probability that an excess is  detectable  at epoch $t$, then we adopt $p(t,\,{\lambda})$ = 1 if $f_{ex}$ $>$ $f_{det}$.  
  If, however, ${\dot M_{dust}}$ is too small for an excess to be detected during steady dust production,  a stochastic 
  mass loss event  can be detected if ${\dot M_{inst}}$ is sufficiently large during a relatively short time. Using the dependence of the excess infrared luminosity  on the dust production rate given by equation (A2),  and averaging over a time long compared to the individual particle decay time but short compared to the stellar evolutionary time, then when $f_{ex}$ $<$ $f_{det}$, we  write that:
 \begin{equation}
p(t,\,{\lambda})\;{\leq}\;\frac{f_{ex}(t,\,{\lambda})}{f_{det}({\lambda})}
\end{equation}
Assuming that $f_{det}({\lambda})$ = 0.02 independent of wavelength and denoting $p_{max}$ as the maximum possible value of $p$,  we show in Figure 6 a plot
 of $<p_{max}({\lambda})>$ between 0.1 Gyr and 4.6 Gyr  for mean dust production  rates given by equations (A4) and (A5).  

If equation (A4) exactly describes the dust production rate,  then, as seen in Figure 5, the excess is never detectable.  If, however, the dust is released in short bursts and equation (A4) only describes the mean loss rate, then as can be seen in Figure 6, there is a small probability that
an excess can be detected.  We can also see from Figure 6 that if equation (A5) describes the dust production rate then during as much as 10\% of the  time  between 0.1 and 4.6 Gyr, a Sun-like star exhibits an infrared excess at 24 ${\mu}$m  produced by asteroid destruction.   
  
  Although there are many poorly known quantities,  Figures 5 and 6 illustrate the
basic result contained within equation (A3).  As long as the time-averaged mass in dust produced by
asteroids is small compared to the time-averaged mass lost from the star as we infer for our solar system,  the expected infrared excess is usually undetectable.

 \newpage
 \begin{center}
 Table 1  -- Helium-Rich White Dwarfs With Low $n$(C)/$n$(Fe)
 \end{center}
 \begin{center}
 \begin{tabular}{llllllll}
 \hline
 \hline
 Star & $t_{cool}$  & ${M}_{Fe}(env)$ & $M_{accrete}$\\
            & (Gyr)       &  (10$^{22}$ g) & (10$^{24}$ g)\\
          \hline
 Ross 640  & 0.7 &  0.08&0.4 \\
 GD 40     & 0.2 &  4 &  6\\
 HS 2253+8023 &   0.2 &  40 & 8  \\
 
 \hline
 \end{tabular}
 \end{center}
 Here, $M_{Fe}(env)$ denotes the mass of iron currently in the convective envelope while $M_{accrete}$ is taken from equation (1).   
  \newpage
\begin{figure}
\plotone{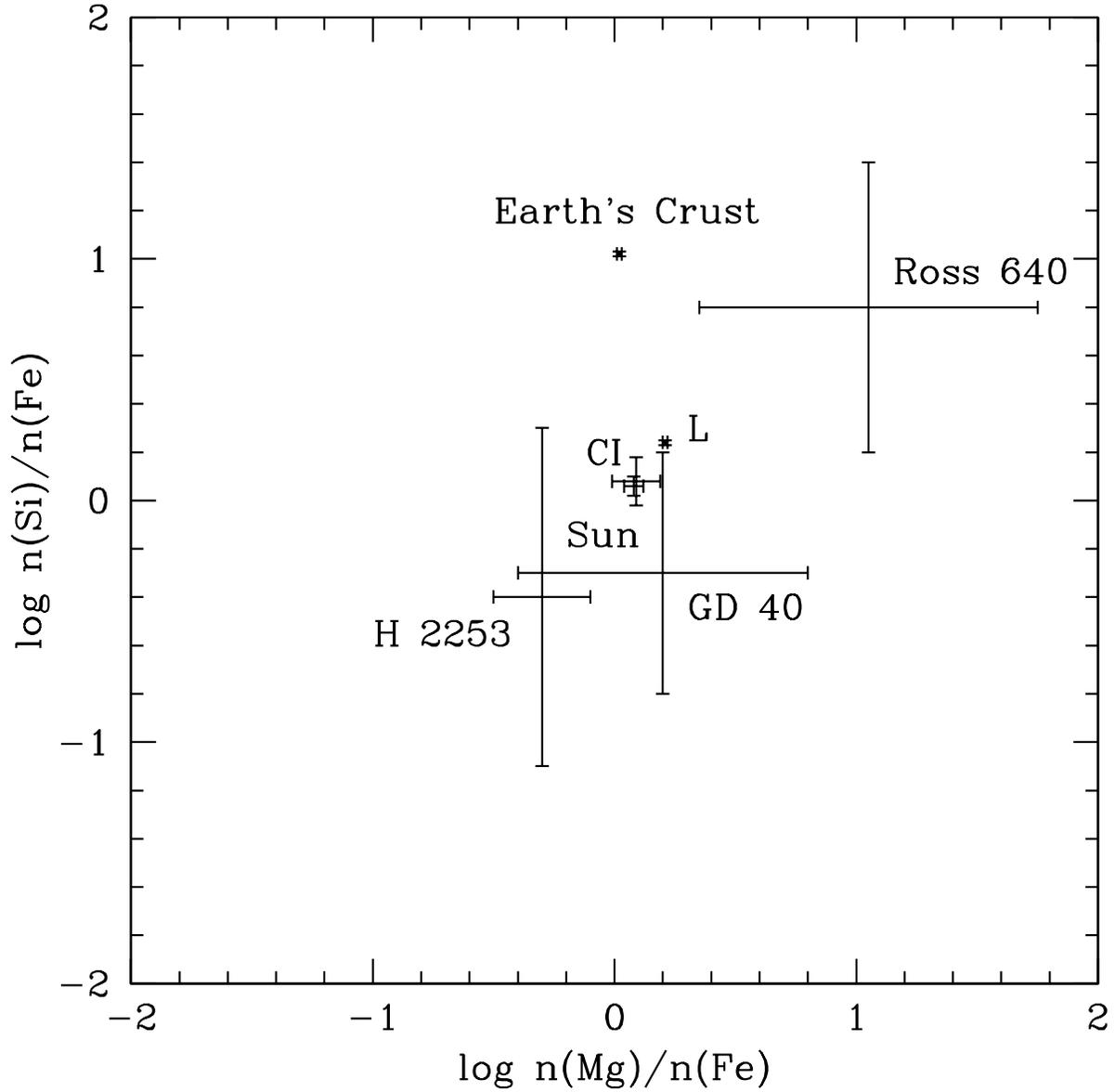}
\caption{Values of $\log$  $n$(Mg)/$n$(Fe) and $\log$ $n$(Si)/$n$(Fe) for the Sun, CI  chondrites (labeled CI), L chondrites (labeled L) the Earth's crust, and three white dwarfs.  We do not display errors bars for the measured abundances in the  L chondrites or the Earth's crust because they are not supplied by the authors of these studies; Nittler et al. (2004) report 
  ${\sim}$20\% abundance variations  among different L chondrites.}
\end{figure}
\begin{figure}
\plotone{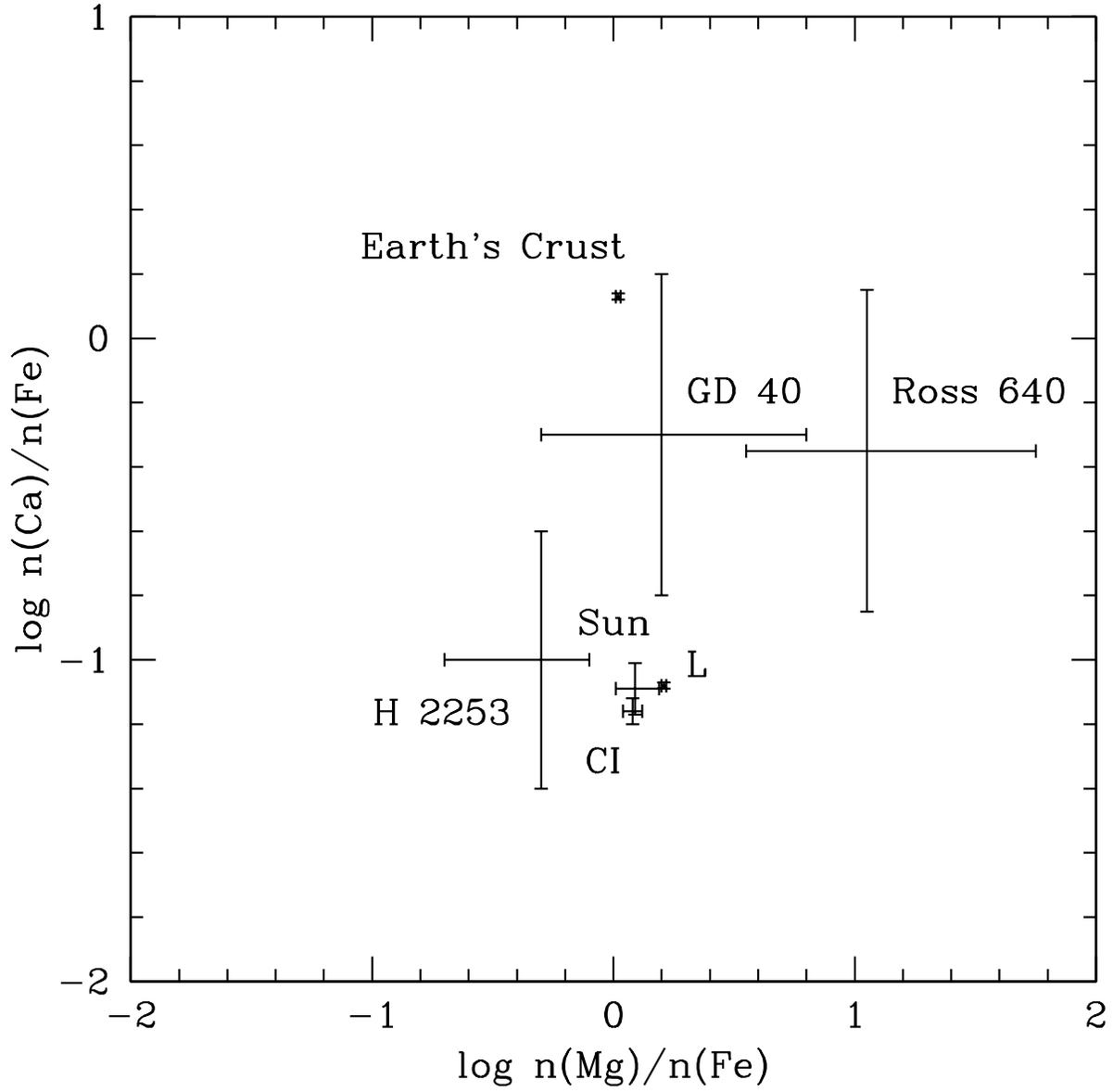}
\caption{Similar to Figure 1 except that  values  of $\log$ $n$(Ca)/$n$(Fe) are compared  with those of $\log$ $n$(Mg)/$n$(Fe).}
\end{figure}
\begin{figure}
\plotone{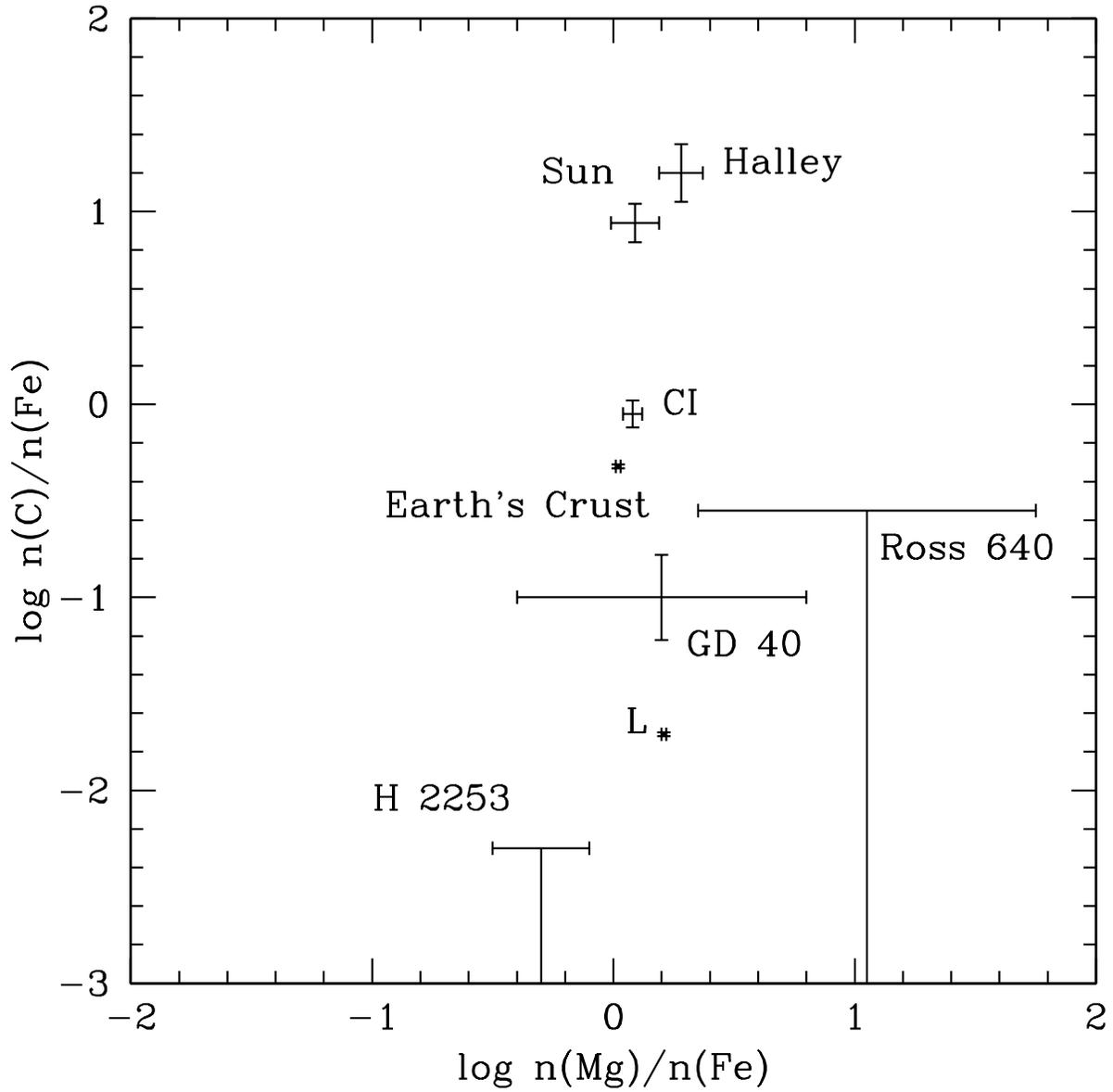}
\caption{Similar to Figure 1 except that  values for $\log$ $n$(C)$n$(Fe) are compared with those of          $\log$ $n$(Mg)$n$(Fe).  We have added a point for the average composition of comet Halley determined by the mass spectrometer on the VEGA-1 spacecraft (Jessberger et al. 1988).}
 \end{figure}
 \begin{figure}
 \plotone{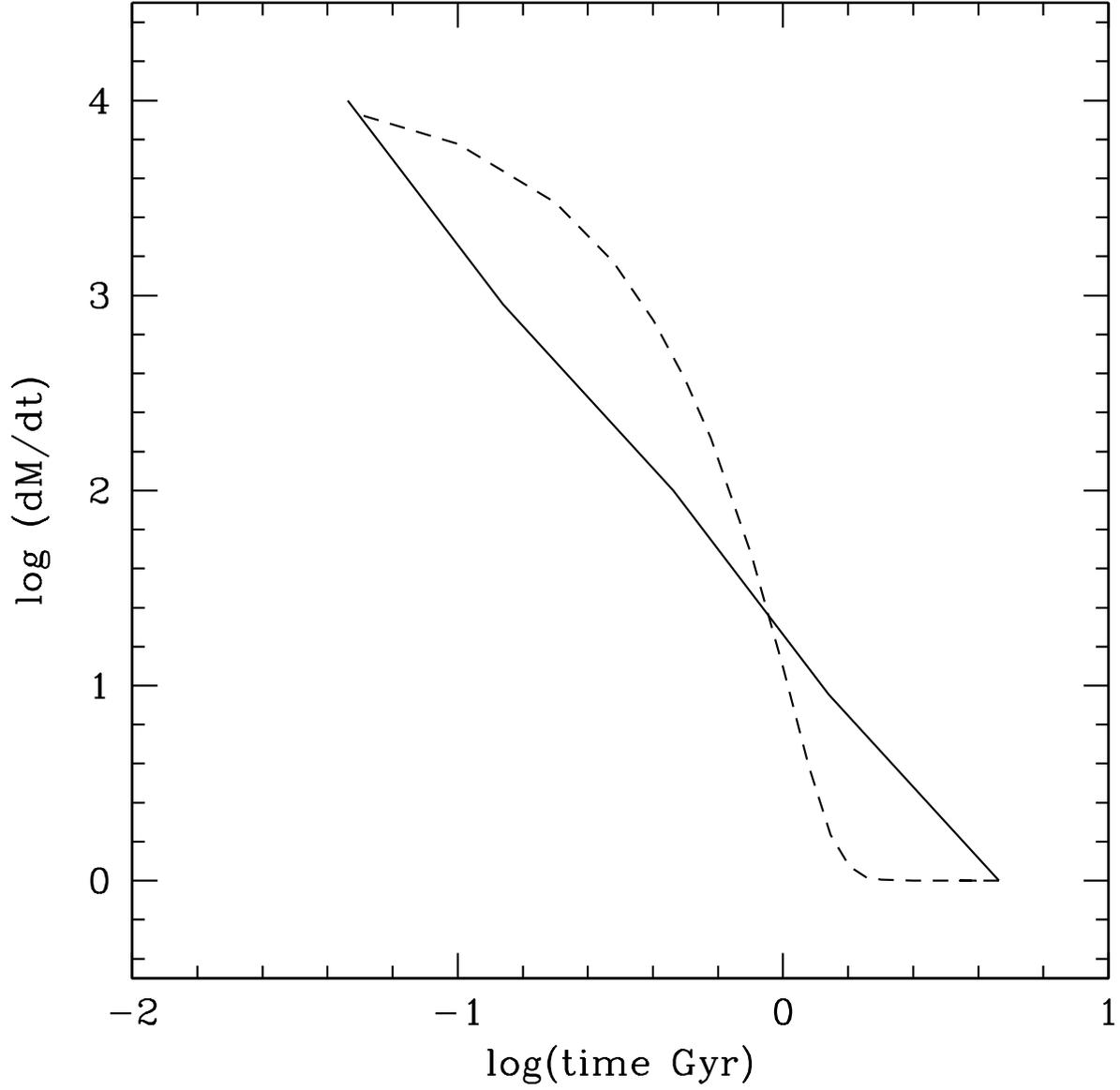}
 \caption{Dust production rate as as a function of time normalized to the current epoch.  The solid and dashed lines are derived from equations (A4) and (A5), respectively.}
 \end{figure}
\begin{figure}
\plotone{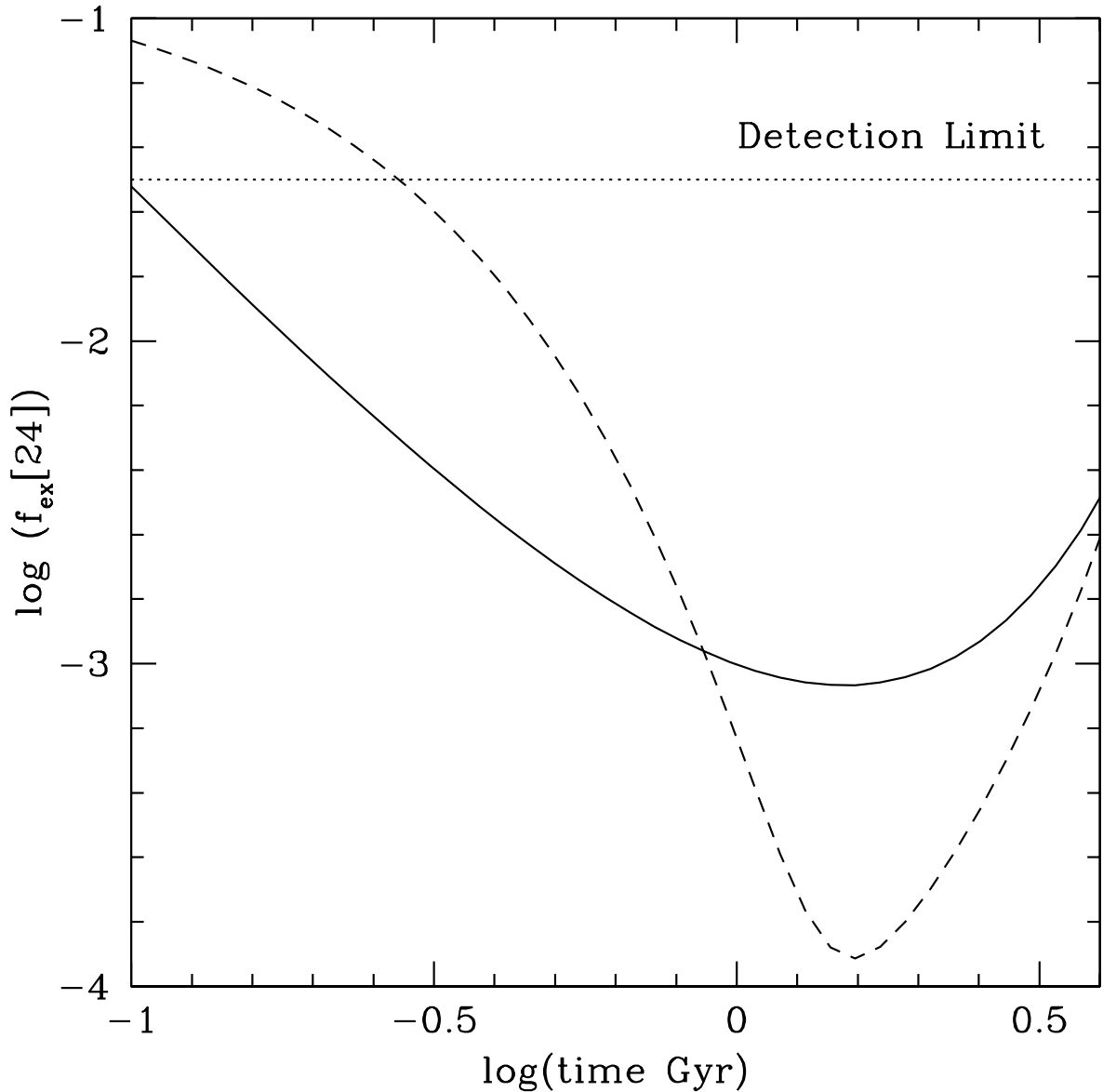}
\caption{Fractional excess at 24 ${\mu}$m as a function of time from equation (A9) for a Sun-like star ($L$ = 1 L$_{\odot}$ and $T_{*}$ = 5780 K)  with the mass loss and dust production  parameters given in the text  shown by the solid line.  The dashed line is the same except that we assume the dust production rate is given by equation (A5) instead of equation (A4).  Infrared excesses may be detectable for $\log$ $f_{exc}[24]$ $>$ -1.5 (Beichman et al. 2006, Bryden et al. 2006, Rieke et al. 2005) }
\end{figure}
\begin{figure}
\plotone{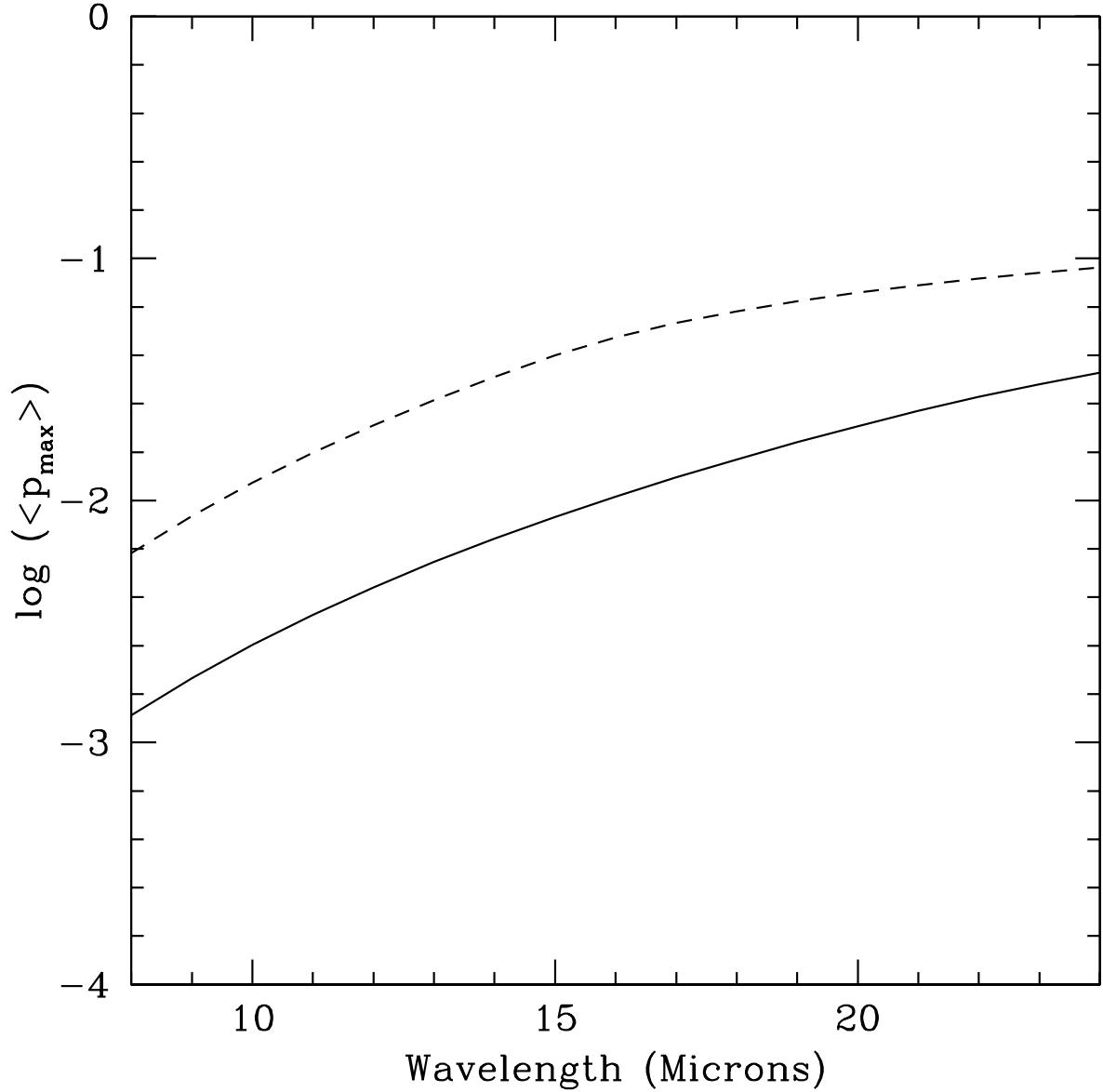}
\caption{Logarithm of the maximum probability of detecting an infrared excess as a function of observing wavelength during the time interval 0.1 Gyr to 4.6 Gyr from equation (A10).  The solid and dashed lines are for mean asteroidal
dust production rates from equations (A4) and (A5), respectively.}
\end{figure}
\end{document}